\documentstyle[twoside,epsfig]{article}

\newcommand{\beq}{\begin{equation}}
\newcommand{\eeq}{\end{equation}}
\newcommand{\beqn}{\begin{eqnarray}}
\newcommand{\eeqn}{\end{eqnarray}}
\newcommand{\bearr}{\begin{array}}
\newcommand{\enarr}{\end{array}}

\newcommand{\toref}[1]{\mbox{(\ref{#1})}}

\newcommand{\eps}{\varepsilon}
\begin{document}

\newcommand {\ee}[1] {\label{#1} \end{equation}}
\newcommand{\be}{\begin{equation}}

\title{Emergence of chaotic behaviour in linearly stable systems}
\author{F. Ginelli\thanks{Dipartimento di Fisica, Universit\`a di Firenze
           and Istituto Nazionale di Fisica della Materia, Unit\`a di Firenze}, 
	R. Livi\thanks{Dipartimento di Fisica, Universit\`a di Firenze
           and Istituto Nazionale di Fisica della Materia, Unit\`a di Firenze}, 
        and A. Politi\thanks{Istituto Nazionale di Ottica Applicata
            and Istituto Nazionale di Fisica della Materia, Unit\`a di Firenze}
        }

\date{\today}

\maketitle

\begin{abstract}
\begin{center}
\parbox{14cm}{
Strong nonlinear effects combined with diffusive coupling may give rise 
to unpredictable evolution in spatially extended deterministic dynamical 
systems even in the presence of a fully negative spectrum of Lyapunov 
exponents. This regime, denoted as ``stable chaos'', has been so far mainly 
characterized by numerical studies. In this manuscript we investigate 
the mechanisms that are at the basis of this form of unpredictable 
evolution generated by a nonlinear information flow through the boundaries.
In order to clarify how linear stability can coexist with nonlinear 
instability, we construct a suitable stochastic model. In the absence
of spatial coupling, the model does not reveal the existence of any
self-sustained chaotic phase. Nevertheless, already this simple regime 
reveals peculiar differences
between the behaviour of finite-size and that of infinitesimal perturbations.
A mean-field analysis of the truly spatially extended case clarifies that the
onset of chaotic behaviour can be traced back to the diffusion process that
tends to shift the growth rate of finite perturbations from the quenched to
the annealed average. The possible characterization of the transition as 
the onset of directed percolation is also briefly discussed as well as
the connections with a synchronization transition.}

\end{center}
\end{abstract} 

\vskip 1.cm
\noindent
\hspace{2.cm} PACS numbers: 05.45+b

\newpage
\section{Introduction}
\label{intro}

Unpredictable evolution in dynamical systems is due to the propagation of 
information.  For instance, the sensitivity of trajectories to infinitesimal 
perturbations implies that any arbitrarily small inaccuracy 
in the determination of the initial conditions is exponentially amplified 
in time, with an average rate associated to the positive component 
of the Lyapunov spectrum. The integral of this component, the so-called 
Kolmogorov-Sinai entropy \cite{ER}, measures the production rate of 
information that flows from the less significant digits of the dynamical 
variables to the more significant ones. In particular, the existence of at 
least one positive Lyapunov exponent is a sufficient condition for 
identifying chaotic dynamics; conversely, a fully negative spectrum is
suggestive of a periodic evolution.

This approach to unpredicatable evolution based on linear stability analysis 
has been developed in the context of finite (low-dimensional) systems. Its 
extension to spatially extended dynamical systems is based on the implcit 
assumption that they can be viewed as a collection of almost independent, 
finite dimensional subsystems.  The existence of a limit Lyapunov spectrum
\cite{K1,LPR} provides strong support to this hypothesis and typical 
chaoticity indicators, like entropies and generalized dimensions, can be 
turned into their corresponding densities \cite{GR1}. In fact, a primary 
interest in the study of space-time chaos is the identification of 
thermodynamic-like properties.

However, Lyapunov instability is not the only source of unpredictability 
in such systems. Actually, information can also flow through the 
boundaries and be transmitted in space by nonlinear mechanisms of front 
propagation.  The so-called chaotic rules of Deterministic Cellular 
Automata (DCA) \cite{Wo} are typical examples of unpredictable evolution in 
the absence of linear instabilities. There, the discreteness of the state
variable prevents the very existence of infinitesimale perturbation, 
while, on the other hand, even isolated ``state-flips'' may propagate through 
the lattice with finite velocity, giving rise to an irregular dynamics. 
In fact, any DCA rule defined over a lattice of $L$ cells is bounded to
exhibit a periodic behaviour, since the number of possible states is finite
($b^L$ if $b$ is the number of possible states in each given site). What makes
a  ``chaotic'' DCA rule different from an ordered one is the exponential 
growth of the recurrence time of the typical configurations. Accordingly, the
unpredictable behaviour is dynamically persistent only in the infinite-size 
limit.  

A very similar unpredictable behaviour, denoted as {\sl stable chaos}, has 
been observed also in coupled map lattice (CML) models \cite{CK,PLOK} of the
type
\begin{equation}
x_i(t+1) = f({\eps\over 2} x_{i-1}(t) + (1 - \eps) x_{i}(t) 
+ {\eps\over 2} x_{i+1}(t) ) \quad ,
\label{cmldyn} 
\end{equation}
where $x_i(t)$ is the continuous state variable at time $t$ on the 
site $i$ of a 1d lattice; the function $f$ is a linearly stable map of the
interval [0, 1] into itself and $\eps$ is the strength of the diffusive 
coupling. It is worth pointing out that this spatial coupling cannot produce 
any linear instability mechanism and the whole spectrum of Lyapunov exponents 
is found to be negative. Accordingly, the CML dynamics must eventually 
approach a periodic stable attractor. This notwithstanding, if $f$ is 
equipped with a sufficiently strong nonlinearity (e.g. a discontinuity or 
a region with rapidly varying slope), one can find a region in parameter 
space, where the ``transient'' evolution towards the periodic attractor
grows exponentially with the system size $L$, analogously to chaotic
DCA rules. Despite there is no rigorous proof of this statement, many 
independent numerical studies confirm such a scenario (see, e.g., 
\cite{LMR,PLOK}). Striking features of this ``transient'' regime are its
stationarity and apparent ergodicity: for instance, space and time correlation 
functions decay exponentially like in usual chaotic phases, while ensemble
avarages coincide with time averages (provided that a size-independent 
pre-transient is discarded). Moreover, the maximum Lyapunov exponent is found 
to approach a stationary negative value, before quite suddenly turning to 
the value corresponding to the eventual attractor. Because of its exponential
growth, the ``transient'' represent the truly relevant regime in the
thermodynamic limit, while the periodic attractor(s) have no practical
significance. 

A careful inspection indicates that the mechanism of information production 
in stable chaos is a flow from the outer (left and right) parts of the chain, 
like in chaotic DCA rules \cite{PLOK}. Accordingly, the unpredictability 
of stable chaos relies on a genuine nonlinear propagation mechanism, 

As we recall in section II, stable chaotic evolution can be detected by 
measuring {\sl damage spreading}, i.e. the average velocity of a front 
propagating into an unperturbed region. This indicator can be viewed as a 
sort of generalization of the standard Lyapunov exponent and the use of a 
proper metric, attributing an increasingly smaller weight to the farther 
sites would make the correspondence more transparent\cite{PLOK}. However,
it is worth noting that the front velocity does not allow defining an analogous 
of a negative Lyapunov exponent, since in the ordered phase a perturbation 
does not only regress but also decreases everywhere in size leading to the 
disappearance of the front itself.

Numerical studies of stable chaos have contributed to shed some light on the 
relationship between this dynamical regime and the appearance of many 
interesting complex phenomena, such as nonequilibrium phase transitions, 
spiral chaos, and the propagation of rough interfaces \cite{KLOP,CLP,KLP}. 
However, little is rigorously known about the underlying mechanisms.
The aim of this paper is precisely to make some progress in this direction
by investigating the conditions under which finite perturbations can propagate 
instead of die out. In analogy to various analytical techniques introduced to
estimate the maximum Lyapunov exponent in standard chaotic CMLs \cite{LPR2,CP},
here we assume that the evolution in phase-space generates a truly random 
pattern, 
characterized by short range correlations and thereby introduce a suitable 
stochastic model describing the evolution in the difference space. Besides
diffusion, the model dynamics allows for a random alternancy of a contraction
and an expansion process the probability of which depends on the perturbation 
size.  First of all, in Sec, II we verify that, with an appropriate
choice of the parameter values, the model is able to reproduce not
only qualitatively but also quantitatively the main features of stable
chaos. Afterwards, we further simplify the rule determining the expansion
process to reduce the subsequent technicalities, while keeping untouched
the key ingredients of the model.

In order to understand how a finite perturbation can be sustained even in the
presence of an average contraction rate, in Sec. III we first discuss 
the uncoupled, i.e. 0-dimensional, case, where, by definition, propagation 
of perturbations is absent. The negative Lyapunov exponent obviously implies
that perturbations are eventually absorbed, so that stable chaos cannot 
exist in this framework. Nonetheless, the presence of a non-uniform 
contraction process yields nontrivial properties of the dynamics. They are 
exemplified by the difference existing between the standard multifractal 
spectrum (associated to infinitesimal perturbations) and the spectrum defined 
in this paper to describe the evolution of finite perturbations. In fact, a 
noteworthy result of Sec. III is that it is possible to define a finite-size 
multifractal spectrum independently of the initial amplitude of the 
perturbation. Finally, in Section III we comment about the connection with 
the finite-size Lyapunov exponents recently proposed by some authors as a 
tool to characterize dynamical unpredictability, beyond standard linear 
stability analysis \cite{Vetal,CT}.

The methods and concepts introduced in Section III are applied in Section IV 
to the study of the spatially extended version of the stochastic model. 
For the sake of space, we limit our discussion to the case of ``democratic'' 
coupling, but it is clear that stable chaos arises in a broad region of
strong coupling.  It is precisely the diffusive coupling to be responsible
for the sustainment of stable chaos. Over small scales, contraction is
more effective than the sporadic amplification: in such circumstances,
diffusion just levels the damping process. Conversely, at larger scales,
diffusion proves to be an effective mechamism to propagate locally
generated amplifications. The net effect is that, in suitable parameter
regions, perturbations self-sustain.

By introducing a factorization hypothesis of the spatial degrees of freedom, 
we obtain a good estimate of the probability distribution associated with the
spatially extended dynamics. In particular, we find a critical value of the
contraction rate separating a dynamical regime, where any perturbation 
eventually vanishes, from a truly chaotic phase, where the average value
of the perturbation remains finite in the infinite-time limit. A mean field 
argument provides a suggestive description of this scenario: the diffusive
coupling induces a shift of the finite-size Lyapunov exponent that grows
from the negative quenched average (corresponding to the uncoupled limit) 
towards the annealed average of the expansion rate. Depending on the
parameter values, this latter average may be strictly larger than 0, thus
implying that the zero-amplitude regime is unstable. 

The transition exibited by the stochastic model is reminiscent of the
``fuzzy'' transition region found in a deterministic CML \cite{CeLP}. 
However, we cannot push the analogy to a quantitative level, since the
stochastic CML dynamics is self-generated and thus it is increasingly regular 
while approaching the transition region (this is the main reason for the 
fuzziness observed in Ref.~\cite{CeLP}), while here the stochastic properties
of the contraction/expansion process is fixed a priori. 
A tight analogy, instead, exists with the stochastic synchronization induced 
by an additive noise \cite{BLT}. Indeed, in this context, the external noise
does not change its robust stochastic features when passing from the
synchronized to non-synchronized regime. It is precisely this analogy which
suggests that the transition described in this paper should belong to the
universality class of directed percolation \cite{DK}. The numerical simulations
described in Sec. V do confirm such an expectation, but subtle problems
still prevent us from still drawing definite conclusions. Such open problems
and the possible future perspectives are summarized in Sec. V.

\begin{figure}[tcb] 
\centerline{\hbox{\psfig 
           {figure=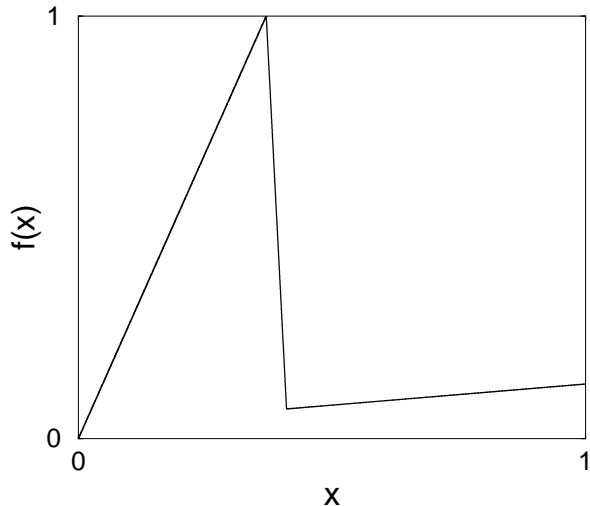,width=7cm,angle=-90}}}  
\caption{ Picture of the local deterministic map \toref{detmap} for typical parameter values.}
\label{localmap} 
\end{figure} 

\section{Generalities} 
\label{gen}

The typical functions used to investigate stable chaos in CML are piecewise 
linear maps of the unit interval of the form (see also Fig.~\toref{localmap}
for a pictorial representation)
\beqn 
f(x)= \left\{  \begin{array}{lr} 
              c_1 x  & 0\leq x < 1/c_1 \\ 
             1-(x-1/c_1)(1-c_2)/\Delta & \quad \quad
                                          1/c_1 \leq x < 1/c_1+\Delta \\
             c_2+c_3[x - (1/c_1+\Delta)] &  1/c_1+\Delta \leq x \leq 1  
\end{array} 
\right.\quad . 
\label{detmap} 
\eeqn 
The most studied case in the literature \cite{PLOK} corresponds to the 
limit $\Delta \to 0$, where $f(x)$ reduces to a discontinuous function.
As we shall comment along this paper, a strong nonlinear component of
map $f(x)$, rather than a true discontinuity, is sufficient to yield 
stable chaos. This is why we prefer to consider here the more general 
case \toref{detmap}.
The parameter range of interest for the present study is when all initial 
conditions (except for a set of zero Lebesgue measure) converge to the 
same periodic orbit. For instance, for $c_1 = 2.7$, $c_2=0.07$, $c_3=0.1$ 
and $\Delta = 0.01$, a stable period-3 orbit
exists with Lyapunov exponent $\lambda = -0.316..$. As already mentioned in
the Introduction, there is a range of values of the diffusive coupling 
$\varepsilon$ defined in (\ref{cmldyn}), where the CML dynamics exhibits
a ``chaotic'' evolution, despite the largest Lyapunov exponent is negative.
Damage spreading analysis provides a first hint about this mechanism, that 
is responsible for the sustainment of irregular behaviour. More precisely, 
while standard chaos amounts to a flow of information from the less to the
more significant digits, stable chaos is generated by a flow of information 
from the outer (left and right) to the inner parts of an infinite chain.
Unfortunately, while the former flow can be ``easily'' studied thanks to the
linearity of the process (in fact, over sufficiently small scales, any smooth 
function can be linearized), the same argument does not apply to the dynamics 
on the left and right edges, that is equally nonlinear at any spatial position.

This is the main reason for the difficulty in deriving the necessary and
sufficient conditions for the propagation of perturbations. Moreover, while
the dynamics of an infinitesimal perturbation can be studied by neglecting
propagation phenomena, the opposite is not possible. One cannot study 
propagation without properly accounting for the local contraction/expansion
mechanisms.

The damage spreading analysis is performed by studying the 
dynamical variable $u_i(t) =|x_i(t)-y_i(t)|$, i.e. the absolute value of
the difference between two test trajectories, $x_i(t)$ and $y_i(t)$, that 
are initially set equal to one another on the right of some lattice site, 
say $i = 0$, while they are assumed to be totally independent on the left. 
With such a setting, the damage spreading analysis amounts to studying the 
propagation of a front separating the region
in space where $ u_i(t) \sim {\cal O}(1)$ (the tail) from the region where
the two trajectories converge to each other, $u_i(t) \sim 0$ (the forefront). 

The first conceptual problem that we have to face is not just the propagation 
of the front, but its self-sustainment in spite of the local average 
contraction rate. In order to shed some light on this crucial point, we have 
simplified the model by assuming that the dynamics is indeed irregular, 
thereby determining whether this assumption is consistent with the sustainment 
of an ${\cal O}(1)$ perturbation. This is analogous to the consistency 
approach developed for the description of standard chaos, where complete 
randomness of the multipliers of the evolution operator in the tangent 
space is assumed in order to estimate the maximum Lyapunov exponent
\cite{LPR2,CP}. 

Accordingly, we introduce a suitable stochastic model to describe the evolution
in the ``difference'' space spanned by $u_i(t)$. Let us start from the simple 
case $\Delta = 0$. With reference to the CML dynamics, the dynamical rule is
composed of two steps. The first one corresponds to the application 
of the standard discrete diffusive operator
\beq 
\tilde{u}_i(t) = (1-\eps)u_i(t)+\frac{\eps}{2}u_{i+1}(t)+ 
\frac{\eps}{2}u_{i-1}(t) \quad , 
\label{diff}
\eeq
where $0 \leq \eps \leq 1$ is the coupling parameter.
The second step contains the stochastic component of the evolution rule,
\beqn
\begin{array}{c} 
u_i(t+1)= \left\{ 
\begin{array}{ll} 
r_i(\tilde{u}_i, t), \quad &  \mbox{w.p.} \quad 
p = \min \left[1, b\tilde{u}_i(t)\right]\\
a\tilde{u}_i(t), & \mbox{w.p.} \quad 1-p 
\end{array}  
\right.\quad,  
\end{array}
\label{disc_stoch2} 
\eeqn 
where, ``w.p.'' is the shorthand notation for ``with probability'', while
$r_i(\tilde{u}_i, t)$ is a random number distributed in the unit interval
according to some probability distribution that depends on $\tilde{u}_i$.

This is the non-trivial part of the stochastic model, defined to mimick the 
evolution of perturbation in the CML dynamics. The first line describes the 
instability mechanism associated with the discontinuity while the second line
describes the contraction of $u_i(t)$ by a constant factor 
\footnote{For the sake of simplicity we assume that the contraction 
rate is a constant, as in \cite{CK}.} 
$ 0 < a < 1$~.  In the CML model, the instability mechanism 
arises whenever the test trajectories, $x_i(t)$ and $y_i(t)$, lie on different 
sides of the discontinuity of the map. In this case, the value taken by the 
difference variable $u_i(t+1)$ is not uniquely determined by the value of 
$u_i(t)$, since it depends also on $x_i(t)$. In particular, for small values
of $u_i(t)$, the instability mechanism occurs quite rarely (it is unlike that
the discontinuity is placed across two nearby trajectories) and $u_i(t+1)$ is 
amplified by a big factor; conversely, for large values of $u_i(t)$, the 
discontinuity plays a role much more frequently, but it is less effective.
The numerical analysis of the CML model shows that the probability $p$ of the 
instability mechanism grows linearly for small values of the difference 
variable $u_i(t)$ and approaches $1$ when $u_i(t)$ does the same. 
Accordingly, in the stochastic model \toref{disc_stoch2}, we have decided to 
schematize this dependence with the simple law $p =\min [1,b\tilde{u}_i(t)]$.

Such a stochastic model can be straightforwardly generalized to cover the case
corresponding to a CML dynamics with a nonzero $\Delta$. The main difference 
is that, whenever $\tilde{u}$ is smaller than $\Delta$, $u$ is expanded by a
fixed factor $\Delta^{-1}$ with a constant probability $b \Delta$, while the 
previous stochastic rule still applies for $\tilde{u} > \Delta$. This amounts 
to assuming that only perturbations larger than $\Delta$ perceive the steep
branch of the map as an effective discontinuity. Accordingly, 
Eq.~\toref{disc_stoch2} is replaced by 
\begin{eqnarray} 
\begin{array}{lc} 
u_i(t+1)= 
\left\{ 
\begin{array}{ll} 
r_i(\tilde{u}_i, t), & \mbox{w.p.} 
\quad p = \min\left[1, b\tilde{u}_i(t)\right] \\ 
a\tilde{u}_i(t), & \mbox{w.p.} \quad 1-p 
\end{array}  
\right.,
& \mbox{if $\tilde{u}_i(t) > \Delta$,}\\ 
u_i(t+1)= 
\left\{ 
\begin{array}{ll}
\tilde{u}_i(t)/\Delta, & 
\mbox{w.p.} \quad p = b\Delta \\
a\tilde{u}_i(t), & 
\mbox{w.p.} \quad 1-p 
\end{array} 
\right.
& \mbox{if $\tilde{u}_i(t) \leq \Delta$.}
\end{array} 
\label{cont_stoch2} 
\end{eqnarray}
One can easily check that these formulae reduce to Eq.~(\ref{disc_stoch2}) 
in the limit $\Delta \to 0$.

The maximum Lyapunov exponent of the CML dynamics corresponds to the time 
average of the expansion rates of infinitesimal perturbations. In the context 
of the stochastic model, it naturally corresponds to the quantity
\beq
\lambda_0 =
\lim_{t \rightarrow \infty} \lim_{u(0) \rightarrow 0} {1\over t}
\sum_{\tau = 0}^{t-1} \ln{||u_i(\tau+1)||\over{||u_i(\tau)||}} =
\lim_{t \rightarrow \infty} \lim_{u(0) \rightarrow 0} {1\over t}
\ln{||u_i(t)||\over{||u_i(0)||}} \quad ,
\label{trueLyapunov}
\eeq
that we still denote as the ``maximum Lyapunov exponent''.

An analytical estimate of $\lambda_0$ can be obtained by a mean-field 
argument, according to which the probability of applying the expansion 
factor $1/\Delta$ is $b\Delta$, while the probability of appling the 
contraction factor $a$ is $1- b\Delta$. One easily obtains
\beq
\lambda_0 \approx b\Delta\ln\left(\frac{1}{\Delta}\right)+(1-b\Delta)\ln a
=\ln a - b\Delta\ln (a\Delta) \quad.
\label{Lyapunov}
\eeq
Numerical simulations indicate that the true value of $\lambda_0$
is generally slightly larger than this mean-field estimate.

Moreover, for $\Delta = 0$, $\lambda_0 \approx \ln a < 0$, while for
increasing values of $\Delta$ it may become positive, indicating that
a standard choatic regime is attained\footnote{This is not entirely correct,
since, as discussed in Ref.~\cite{PT}, the standard chaotic regime occurs 
when a stronger constraint is met: the linear propagation velocity coincides
with the nonlinear one.}. Here, we are interested in studying only the 
parameter region where $\lambda_0$ remains negative.

\subsection{Comparison between stochastic and deterministic models}

The reliability of the stochastic models (\ref{disc_stoch2})
and (\ref{cont_stoch2}) has been tested by numerical
simulations which show that both rules (\ref{disc_stoch2}) and 
(\ref{cont_stoch2}) together with the diffusive coupling \toref{diff}, 
exhibit the same qualitative features of the CML dynamics (\ref{cmldyn}),
(\ref{detmap}). In particular, when the contraction is relatively strong, 
any initial perturbation $u_i(0)$ is quickly absorbed to the fixed point 
$u_i=0 \quad \forall i$: this regime corresponds to the ``ordered'' or, 
equivalently, to the ``synchronized'' phase. On the other hand, for weaker
contractions, almost any initial condition evolves towards an irregular 
spatial structure like the one shown in Fig.~\ref{singlefront}. In this case,
the space average 
\beq
\bar{u}(t)=\frac{1}{L}\sum_{i=1}^L u_i(t) \quad 
\label{space_av}
\eeq
remains finite and $\bar{u}(t)= u^*$, independently of $t$ and $L$ for 
sufficiently large sizes. To be more precise, perturbations of the order 
${\cal O}(1)$ survive only for a finite time $\tau$ also in the chaotic phase 
of the stochastic model. Nevertheless, in analogy to the CML model, $\tau$
grows exponentially with the lattice size $L$. Furthermore, ensemble averages 
indicate that this regime is asymptotically stationary in time and the 
comparison with time averages (obviously performed over times
much shorter than $\tau$) indicate that ergodicity holds as well.
It is therefore meaningful to define the single-site probability distribution 
$Q(t,v)$ of finding a perturbation $u_i$ in between $v$ and $v+dv$ at time 
$t$, 
and its stationary limit ${Q}(v)$, attained for large enough times. It is also 
worth introducing the space and ensemble averages 
$m(t)=\langle \bar{u}(t) \rangle$ (from here on, unless otherwise stated, 
$\langle \cdot \rangle$ denotes an ensemble average) and the stationary limit, 
$m(t) \to \tilde{m}$.

\begin{figure}[tcb] 
\centerline{\hbox{\psfig 
           {figure=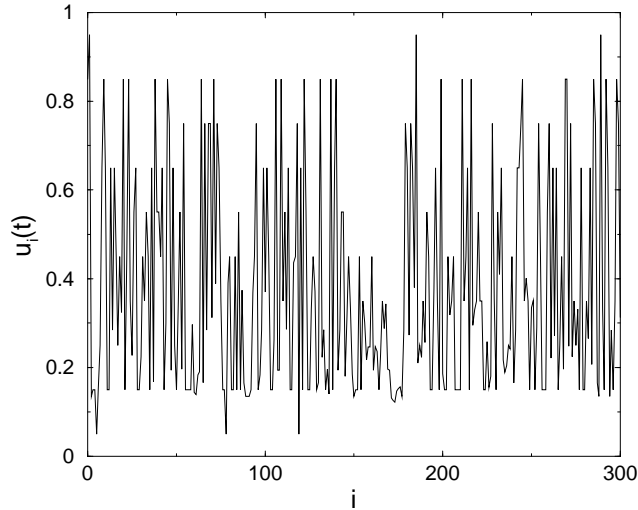,width=7cm,angle=-90}}}  
\caption{Snapshot of the perturbation profile in the chaotic phase 
of the stochastic model (\ref{diff}, \ref{cont_stoch2}) at time $t=500$.}
\label{singlefront} 
\end{figure} 

In the numerical investigations we have assumed no-flux boundary conditions 
as they preserve the $u=0$ fixed point and are reasonably harmless in the 
chaotic phase. The $\tilde{u}$-dependent probability distribution of 
the random variable $r_i(\tilde{u}_i,t)$ has been reconstructed from the CML
dynamics. Moreover, in order to avoid boundary effects in the damage spreading
analysis, the lattice size has always been chosen in such a way that the 
perturbation front never reaches the lattice edges during each simulation.

\begin{figure}[ctb] 
\centerline{\hbox{\psfig {figure=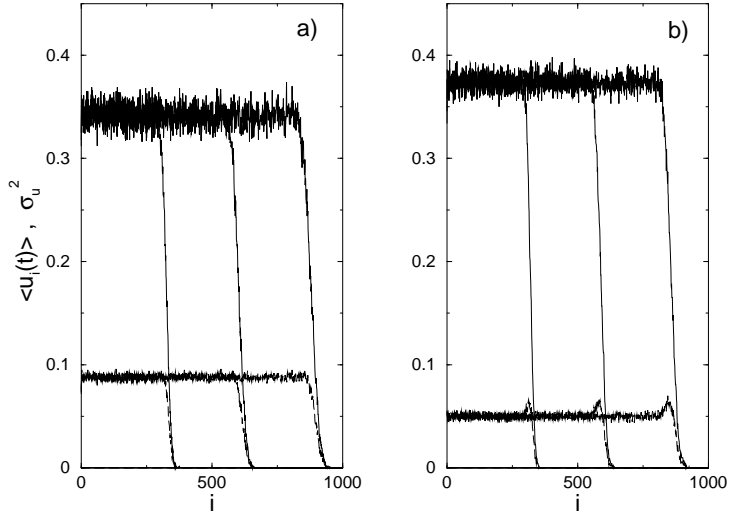,width=7cm,angle=-90}}}  
\caption{ Average front profiles (upper curves) and the corresponding variances
(lower curves) at three different times, $t=500$, $t=1000$ and $t=1500$. Panel 
(a) refers to the deterministic CML model (\ref{cmldyn}),(\ref{detmap}), with 
parameter values $c_1= 2.7$, $c_2=0.07$, $c_3=0.1$, $\Delta=0$, $\eps=2/3$; 
averages have been performed over $10^3$ initial conditions. Panel (b) refers 
to the stochastic model (\ref{diff}), (\ref{disc_stoch2}), with parameter 
values $a=0.9$, $b=1.7$, $\Delta=0$, $\eps=2/3$; averages have been performed 
over $10^3$ realizations, while the probability distribution for 
$r_i({\bar u}_i,t)$ and the values of parameters $a$ and $b$ have been 
determined from the CML model depicted in the left panel.} 
\label{frontprofile} 
\end{figure} 

In order to investigate damage spreading phenomena, the initial conditions 
have been fixed by imposing $u_i(0) = 0$ for $i>0$ and 
randomly choosing $u_i(0)$ with a uniform probability distribution in the 
interval $[0,1]$ for $i \le 0$. The statistical fluctuations showed by 
$u_i(t)$ have been smoothed out by performing ensemble averages (over different 
realizations of the stochastic process) of the spatial
configurations at equal times. In the chaotic phase, the initial ``kink''-like
structure persists: the front connecting the perturbed with the synchronized
region moves with a fluctuating velocity. The ensemble averages corresponding
to three different times are reported in Fig.~\ref{frontprofile}b, where they
are compared with the results directly obtained from the CML model (for the 
sake of space we limit ourselves to consider the case $\Delta = 0$a), 
where ensemble averages are performed over different initial conditions.
Several observations are in order. First, we notice that, in spite of the
simplifications introduced in the stochastic model (besides the lack of
space-time correlations, we have indeed assumed a constant contraction rate
as if the two branches of the local map \toref{detmap} had the same slope),
there is a reasonable agreement with the CML data. In particular, it can be
seen that the average height $u^*$ is definitely smaller than 1 in both cases:
the reason can be traced back to the combined effect of the contraction 
mechanism with the diffusive process. Somehow larger differences can be
observed in the behaviour of the variance 
$\sigma_u^2 = \langle (u_i(t))^2 \rangle - \langle u_i(t)\rangle ^2$ but
they can be attributed more to the approximation in the description of
the perturbed region, rather than to peculiarities of the propagation.

In fact, if we look at the width $w(t)$ of the average front, defined as the 
distance between the rightmost sites where $\langle u_i(t) \rangle$ is 
larger than $0.8 u^*$ and, respectively, $0.2 u^*$, we observe a nearly
square-root growth in both the CML and the stochastic model (see 
Fig.~\ref{frontamplitude}, where a nearly quantitative agreement is also
observed). Moreover, we have verified that, keeping the contraction
parameter $a$ fixed, and increasing $\Delta$ up to small values 
(e.g. $\Delta \sim 0.01$), $v_F$ slightly increases in both models.

\begin{figure}[ctb] 
\centerline{\hbox{\psfig {figure=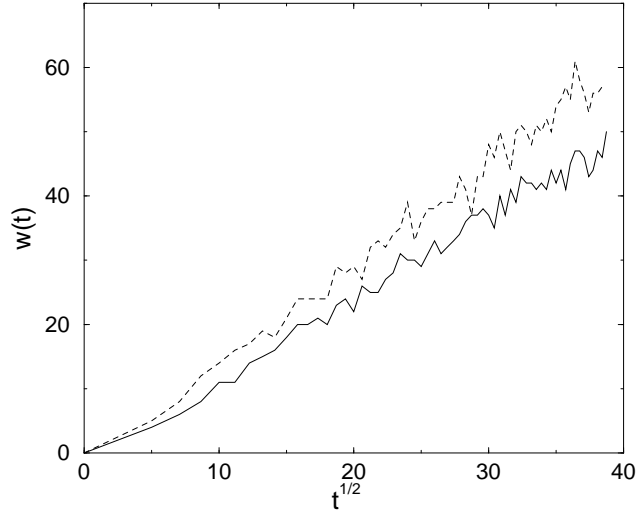,width=7cm,angle=-90}}} 
\caption{The average front size $w(t)$ of the stochastic (full line) and CML 
(dashed line) models described in the caption of Fig.~\ref{frontprofile}
plotted versus the square root of time. See the text for the definition of
$w(t)$.}
\label{frontamplitude} 
\end{figure} 

On the basis of the results reported in Fig.~\ref{frontamplitude}, one can
summarize our observations of the front dynamics by effectively assuming a 
stepwise shape for the profile at any time and approximating its motion with 
a diffusive process with drift. More precisely, we can write
\begin{eqnarray}
\langle u_i(t)\rangle &\approx& \frac{u^*}{\sqrt{2\pi D t}}
\int_{0}^{+\infty} \Theta [x-(i-v_Ft)]\,\exp \left(-\frac{x^2}{2Dt}\right)dx
 =\nonumber\\
 &=& \frac{u^*}{2}\left[ 1-\mbox{erf}\left(\frac{i-v_Ft}
{\sqrt{2Dt}} \right)  \right],
\end{eqnarray} 
where $v_F$ is the average front velocity, while $D$ is the ``diffusion''
coefficient accounting for the square-root growth in time for the standard
deviation of the front position $x$ (assumed to be continuous, for the sake 
of simplicity). 

\subsection{Further simplifications of the stochastic models}
As we have verified that the qualitative features of the front dynamics
do not depend on the shape of the probability distribution of 
$r_i(\tilde{u}_i,t)$, we have decided to simplify the stochastic models by 
assuming a $\delta$-like distribution, i.e. that $r_i(\tilde{u}_i,t)=1$.
Moreover, in order to get rid of unnecessary technical complications we 
shall assume that $b=a$. Notice also that, having chosen  
$r_i(\tilde{u}_i,t) = 1$, we are obliged to assume $b<1$, in order to avoid 
the appearence of a fictitious fixed point $u^* = 1$. From now on we study 
the following model 
\begin{eqnarray} 
\tilde{u}_i(t) = (1-\eps)u_i(t)+\frac{\eps}{2}u_{i+1}(t)
+ \frac{\eps}{2}u_{i-1}(t),\nonumber\\ 
\begin{array}{lc} 
u_i(t+1)= 
\left\{ 
\begin{array}{ll} 
1, & \mbox{w.p.} \quad p=a\tilde{u}_i(t) \\ 
a\tilde{u}_i(t), & \mbox{w.p.} \quad 1-p
\end{array}  
\right.,
& \mbox{if $\tilde{u}_i(t) > \Delta$,}\\ 
u_i(t+1)= 
\left\{ 
\begin{array}{ll}
\tilde{u}_i(t)/\Delta, & \mbox{w.p.} \quad p=a\Delta \\
a\tilde{u}_i(t), & \mbox{w.p.} \quad 1-p
\end{array} 
\right.
& \mbox{if $\tilde{u}_i(t) \leq \Delta$,}
\end{array} 
\label{cont_stoch} 
\end{eqnarray}
that we call {\it Continuous Stochastic Model} (CSM). In the limit 
$\Delta \to 0$ it will be named {\it Discontinuous Stochastic Model} (DSM).

The only parameter that we are going to consider in the following sections is
the contraction parameter $a$: this is sufficient to identify and
characterize the relevant transition from the chaotic/unsynchronized phase
to the ordered/synchronized one.

\section{The zero dimensional case}
\label{0dim_model}

The most important problem of linearly stable chaos is to understand how finite 
perturbations can propagate in spite of the average local contraction.
To clarify this point, in this section we consider the zero-coupling limit, 
i.e. the 0-dimensional case. We shall see that even if ``chaotic'' motion
cannot be sustained whenever the Lyapunov exponent is negative, the lack of 
a uniform contraction induces anyhow non-trivial properties.

Our first observation concerns the probability $P(t,u(0))$ for a 
finite-amplitude perturbation $u(0)$ to be never amplified by the instability 
mechanism over a time $t$. In the DSM, such a probability can be easily 
factorized as
\begin{eqnarray}
P(t, u(0)) =\prod_{n=1}^{t} (1-u(0)a^n) = 
   \exp \left[\sum_{n=1}^{t}\ln (1-u(0)a^n)\right] =\nonumber\\ 
   =\exp \left[ -\sum_{k=1}^{\infty}\frac{u(0)^k}{k}\sum_{n=1}^{t}a^n\right] 
    =\exp \left[ -\sum_{k=1}^{\infty}\frac{u(0)^k}{k}\frac{a^k(1-a^{kt})}
   {1-a^k}\right]
   \nonumber\\ 
  \stackrel{t\rightarrow \infty}{\longrightarrow} \exp 
 \left[ -\sum_{k=1}^{\infty}\frac{(u(0)a)^k}{k(1-a^k)}\right] := \tilde{P}(u(0))\quad.
 \quad\quad
\label{P0} 
\end{eqnarray}
One can see that, for $t\to\infty$, $P(t, u(0))$ approaches a finite value 
$\tilde{P}(u(0))$ that is both strictly larger than 0 and smaller than 1 
for any value of $u(0)$ (if $0<a<1$). The same conclusion can be drawn also 
for the CSM, although the algebra is more complicate in that case. The 
inequality $\tilde{P}(u(0))>0$ 
indicates that the occasional amplifications are not so strong as to prevent 
the eventual absorption of the perturbation (this is consistent 
with our goal to deal with linearly stable processes). On the other hand, 
the inequality $\tilde{P}(u(0))<1$ indicates that the amplification process 
cannot be neglected.  Notice that, since this holds true independently of 
$u(0)$ (although $\tilde{P}(u(0)) \to 1$ for $u(0) \to 0$), there is a 
difference with the convergence to a stable fixed point in a topologically 
chaotic map (think, e.g., of the logistic map in one of the stability windows
that follow the first period doubling) since, in that case, there would be 
a threshold (corresponding to the border of the basin of attraction), below 
which a monotonous contraction would start.
A closer similarity exists with the so-called strange nonchaotic attractors, 
as they are characterized by a nonmonotonous contraction even arbitrarily 
close to the attractor \cite{SNA1,SNA2}. 

The multifractal theory \cite{BS} provides the most appropriate framework to 
characterize this system. By following this approach, devised with
reference to infinitesimal perturbations, we introduce the exponential growth 
rate of a finite initial perturbation $u(0)$,
\beq
\Lambda(t, u(0))=\frac{1}{t}\ln\left(\frac{u(t)}{u(0)}\right) .
\label{EFSLE}
\eeq
In the limit $u(0)\rightarrow 0$, $\Lambda(t, u(0)) \to \lambda(t)$, the 
equivalent in this context of the usual finite-time Lyapunov exponent 
\cite{FTL}. The proper indicator to look at is the
probability distribution ${\cal P}(\Lambda, t,u(0))$ to find a growth 
rate between
$\lambda$ and $\Lambda + d\Lambda$ at time $t$ starting from $u(0)$.
More precisely, we introduce the finite-size multifractal spectrum
\beq
H(\Lambda)=\lim_{t \rightarrow \infty}
\left[\frac{1}{t}\ln {\cal P}(\Lambda, t,u(0))\right].
\label{multifractal2}
\eeq
As it will become clear later, $H(\Lambda)$ is independent of $u(0)$ (as long
as $u(0) >0$)~. Nevertheless, we shall show that it differs 
from the standard multifractal spectrum $h(\lambda)$, obtained by taking 
the limit $u(0) \to 0$ before the infinite-time limit
\beq
h(\lambda)= \lim_{t \rightarrow \infty} \lim_{u(0) \rightarrow 0}
\left[\frac{1}{t}\ln {\cal P}(\Lambda, t,u(0))\right].
\label{multifractal1}
\eeq
In other words, the order of the two limits is crucial for understanding 
the difference between the behaviour of infinitesimal and finite 
perturbations.  In the following, we shall compare the 
multifractal distribution of $\lambda(t)$ ({\it linear analysis}) with
that of $\lambda(t,u(0))$ ({\it nonlinear analysis}) in both stochastic models. 

\subsection{Linear analysis}
\label{linan}
We start from the CSM, as the DSM is nothing but a limit case of the former 
one. The linear approximation amounts to assuming $u_i(t) < \Delta \quad 
\forall i, t$.  In this case, standard combinatorial analysis implies 
\begin{eqnarray}
h_{\Delta}(\lambda) &=& 
\frac{\ln(\Delta) + \lambda}{\ln(a\Delta)}\left[\ln(1-a\Delta)-
\ln\left(\frac{\ln(\Delta) + \lambda}{\ln(a\Delta)}\right)\right]\nonumber\\
&&+\frac{\ln(a)-\lambda}{\ln(a\Delta)}\ln\left[\frac{a\Delta
\ln(a\Delta)}{\ln(a)-\lambda}\right],
\label{cont_linear}
\end{eqnarray}
where we have made explicit the dependence on the parameter $\Delta$.
In the limit $\Delta \to 0$, the above expression reduces to
\beq
h_0(\lambda) = \lim_{t\rightarrow \infty}\frac{1}{t}\ln {\cal P}[\lambda(t)]
\stackrel{\Delta \rightarrow 0}{\longrightarrow} \ln(a)-\lambda, \quad
\lambda > \ln a \quad .
\label{disc_linear1}
\eeq
On the other hand, performing the limit $\Delta\rightarrow 0$ before the 
$t \to \infty$ limit would yield the trivial result $h_0(\lambda) =0$ with the 
support of $h_0$ restricted to the point $ \lambda = \ln a$. Thus, the 
non-commutativity of the limits $\Delta \to 0$ and $t \to \infty$ reveals
that the discontinuous case is a singular limit of the CSM. In other words,
the multifractal spectrum of the discontinuous model depends on the way it
is defined. We prefer to adopt a ``physical'' point of view, i.e. to consider 
the discontinuity as the limit of a negligible $\Delta$, which
corresponds to taking first the $t\to\infty$ limit.

\subsection{Nonlinear analysis}

We now consider a finite perturbation $u(0)$ in the simple context $\Delta =0$.
From Eq.~\toref{EFSLE} we see that $\Lambda$ is uniquely determined from
the knowledge of $u(t)$ and $u(0)$ (and, obviously, the time $t$).
Accordingly, the knowledge of ${\cal P}(\Lambda,t,u(0))$ is fully equivalent 
to that of the single-site probability distribution $Q(t,u)$, together with
the initial condition
\beq
Q(0, u)=\delta(u-u(0)) \quad.
\label{t0value}
\eeq
It proves useful to introduce the notation $u=a^n$ and $u(0)=a^{n_0}$, where 
$n$ and $n_0$ are real variables $\geq 0$. Eq.~\toref{EFSLE} can be rewritten as
\beq
\lambda=\lambda(t,n, n_0)=\frac{1}{t}\ln\left(
\frac{a^n}{a^{n_0}}\right)=
\frac{n-n_0}{t}\ln a \quad.
\label{lambdafin}
\eeq
Once $Q(t, n)$ is known, Eq.~\toref{lambdafin} allows one reconstructing the 
corresponding probability distribution ${\cal P}(\Lambda, t, u(0))$. Two 
possibilities are in order, either the system has never been ``kicked'', i.e. 
reset to 1 by the instability mechanism, in which case the initial value has 
been contracted $t$-times by a factor $a$, or it has received at least one 
kick, loosing memory of the initial condition. In the former case, occurring 
with probability $P(t, u(0))$, $\Lambda(t, u(0))=\ln a$ or, equivalently, 
$n=n_0+t$. In the latter case, $\Lambda(t, u(0)) > \ln a$
and the accessible values of $n$ are restricted to positive integer numbers
strictly smaller than $n_0+t$ (since that the maximum possible contraction 
factor in $t$ time steps is $a^t$). Accordingly, $n$ can be interpreted as the
elapsed time since the last kick, and the probability distribution $Q(t,u)$ can 
be factorized as the product of the probability $G(t-n, n_0)$ of receiving a 
kick at time $t-n$ for an initial perturbation $u(0)=a^{n_0}$ by the 
probability $P(n,1)$ of not being kicked anymore for the remaining $n$ 
time steps,
\beq
Q(t,n)=G(t-n,n_0)P(n, 1)
\label{pippo}
\eeq
(in the remaining part of this section and in App. A, with no ambiguity
arising, we denote with $Q$ also the probability density of the logarithmic 
variable $n$). Moreover, we impose the condition $P(0, u(0)):=1$ to extend the
validity of Eq.~\toref{pippo} to the case $n=0$.

The probability $G(t, n_0)$ can be recursively expressed as the probability 
of receiving the very first kick at time $t$ plus the probability of 
receiving the second last kick at any previous time, i.e.
\beq
G(t,n_0)=a^{t+n_0}\prod_{k=1}^{t-1}(1-a^{k+n_0})
+\sum_{k=1}^{t-1}G(k, n_0)a^{t-k}P(t-1-k, 1),
\label{pippo3}
\eeq
with $G(1, n_0):=a^{n_0+1}$. Eq.~\toref{pippo3} can be numerically iterated 
in the large time limit to obtain the expression of the multifractal 
distribution \toref{multifractal2}. The analysis performed in App.~\ref{appA} 
shows that $H(\Lambda)$, defined as in \toref{multifractal2}, is a segment of 
straight line restricted to the open interval $(\ln a, 0)$ of negative values. 
The slope of this straight line depends on the contraction parameter $a$ but is
independent of $u(0)$. In fact, in the appendix we show that the slope can 
be obtained by solving an eigenvalue problem, where $u(0)$ enters to specify 
the initial condition but not the operator itself. An approximate
analytic solution is also determined, which confirms the numerical
observation that the slope increases monotonously from -1 
(for $a\rightarrow 0$) to 0 (for $a\rightarrow 1$). Curves 2 and 4 in 
Fig.~\ref{0dim_curves} correspond to the nonlinear and linear analysis 
of the discontinuous case for $a=0.7$, respectively. 
The latter 
curve lies well below the former one, indicating that the linear analysis 
leads to an underestimation of the fluctuations. This is a general fact 
holding for all values of $a$ in the meaningful range $[0,1]$. The difference  
must be attributed to the sporadic amplifications due to the discontinuity:
it is remarkable that the finite-size spectrum is independent of the 
initial condition.

It is now important to test whether the difference between linear and
nonlinear curves persists also when the discontinuity is removed, i.e. 
in the CSM. For the sake of simplicity, we fix $\Delta$ equal to $a^{\bar{n}}$ 
and suppose that both $\bar{n}$ and $n_0$ are non-negative integers, 
so that $u(t)$ is defined on a discrete subset 
of the unit interval, $\{a^n\}_{n=0,1,\ldots}\;$. This assumption does not 
affect the main conclusions while it allows one writing a simple recursive equation
for the probability distribution $Q(t, n)$:
\begin{eqnarray}
Q(t+1,0)&=&
\sum_{k=0}^{\bar{n}}Q(t,k)\,a^{k+1}, \nonumber \\
Q(t+1,n)&=&\left\{
\begin{array}{lc}
(1-a^n)\,Q(t,n-1)+a^{\bar{n}+1}\,Q(t, n+\bar{n}) & 0<n\leq\bar{n}\\
(1-a^{\bar{n}+1})\,Q(t, n-1)+a^{\bar{n}+1}\,Q(t, n+\bar{n}) & n>\bar{n}
\end{array}
\right.
\label{uncleduck}
\end{eqnarray}
Equations \toref{uncleduck}, with the initial conditions $Q(0,n)
=\delta_{n,n_0}$
and boundary conditions $Q(t,n)=~0$ $\forall\, t > n+n_0$, can be 
numerically iterated to obtain the finite-size multifractal distribution 
in the continuous case. 

\begin{figure}[bct] 
\centerline{\hbox{\psfig {figure=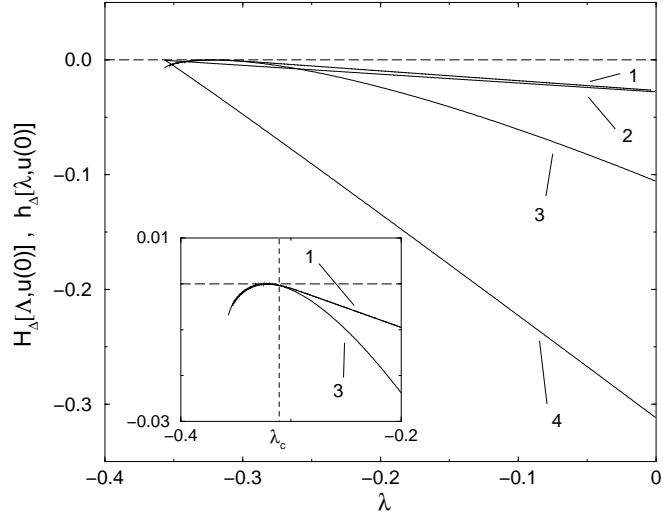,width=7cm,angle=-90}}}
\caption{Probability distributions of the exponential growth rate 
\toref{EFSLE} in the large time limit for the stochastoc model \toref{cont_stoch} in
the 0-dimensional case, with $a = 0.7$.
Lines 1 and 3 refer to the continuous case 
($\Delta=0.0097$), with a finite size or an infinitesimal initial 
perturbation, respectively, while lines 2 and 4 refer to the 
discontinuous case 
($\Delta\rightarrow 0$), still with a finite size or infinitesimal 
initial perturbation, respectively. 
Deviation from straight line of curve 2 is due to finite time effects.
In the inset the difference between curves 1 and 3 is magnified.}
\label{0dim_curves} 
\end{figure} 

Since the development of analytical techniques to determine $H_\Delta$ is by 
far more complex than in the previous case, we have limited ourselves to
determine the multifractal spectrum numerically. The linear and nonlinear 
spectra are reported in Fig.~\ref{0dim_curves} for $\Delta = 0.0097$ 
(curves 3 and 1, respectively). Let us first notice that the finite-size
spectrum is again independent of the initial condition $u(0)$ and this makes
$H_\Delta$ a well defined quantity: the effect of $u(0)$ is just to affect
the convergence to the asymptotic spectrum. Furthermore, we can see that
the nonlinear curve lies well above the standard multifractal spectrum,
indicating that it is not the discontinuity to be responsible for the 
difference between the behaviour of infinitesimal and finite-size
perturbations. Moreover, the overall closeness of curves 1 and 2 reveals that 
the removal of the discontinuity does not introduce significant differences
in the finite-size spectrum.

Finally, let us closely compare $H_\Delta(\Lambda)$ with 
$h_\Delta(\lambda)$: the inset in
Fig.~\ref{0dim_curves} reveals that the two coincide for $\Lambda, \lambda <\lambda_c$
(see the dashed line), while above some critical value $\lambda_c$ 
the finite-size spectrum
continues as a straight line, while $h_\Delta$ decreases faster. In the
language of thermodynamic formalism, the linear behaviour of $H_\Delta$
is suggestive of a phase transition, from small values of $\Lambda$, 
that are correctly 
described by the linear analysis, to large values of $\Lambda$, 
where the finite
character of the perturbation cannot be neglected. This point would certainly
require a more detailed analysis to provide a more solid background to the
above arguments, but we avoid this as it would drive us too far from what is
the main goal of the present paper. 

The analysis performed in this section has allowed us introducing well
defined observables to deal with finite-size perturbations. We cannot, however,
avoid commenting on an alternative class of tools that have been devised to
deal with this problem, although mainly in the context of truly chaotic
systems. Since in realistic physical conditions perturbations are always 
finite, it is very tempting to introduce a growth rate to characterize
perturbations of different sizes. However, serious conceptual problems
are immediately encountered if one tries to define truly finite-size Lyapunov 
exponents. On the one hand, the finiteness of the size induces a dependence
on the norm utilized, on the other hand it requires that the Lyapunov
exponent has to be defined for a finite-time resolution, since perturbations 
change size over time. This latter implication is rather crucial in that so
defined Lyapunov exponents are not self-averaging quantities 
(see Ref.~\cite{LK} for a detailed discussion of the problem).

In spite of such limitations, finite-size Lyapunov exponents may carry
useful information, although one has to be careful in interpreting them. 
Let us, for instance, look at
\beq
\Lambda_0(u) = \left \langle \ln \frac{u'}{u} \right \rangle
\eeq 
where $u'$ is the first iterate of $u$ in the DSM and the average is performed
over all possible realizations of the stochastic process. Simple algebra yields
\beq
\Lambda_0(u) = \ln a - au \ln (a u)\quad.
\eeq
In the limit $u \to 0$, $\Lambda_0(u)$ reduces correctly to the true Lyapunov
exponent $\log a$. If $a$ is not too small ($a > 1/e$, where $e$ is the Neper
number), $\Lambda_0(u)$ has a maximum value for an intermediate value of $u$ 
and, more interestingly, the maximum value of $\Lambda$ is larger than 0, 
if $a > \exp(-1/e) = 0.692..$. Therefore, this result is suggestive of a phase
transition from a regime (small $a$) where perturbations of all sizes
decrease, to a regime where sufficiently large pertubations expand (and can,
in principle, self-sustain). However, we already know that this conclusion is
incorrect: the reason is precisely that $\Lambda_0(u)$ is an average quantity, 
and fluctuations must be taken into account. In fact, a different scenario
arises, if we define the contraction rate by taking the logarithm of the
average expansion factor. In this case, we obtain
\beq
\Lambda_1(u) = \ln a + \ln (2 - au),
\label{fsle}
\eeq
an expression that can be larger than zero even at $u=0$, where it actually
attains its maximum value! In practice, one has to be very careful in
drawing meaningful conclusions from any expression of the finite-size
Lyapunov exponent. In particular, since $u$ can never be larger than 1, while 
it can become arbitrarily small, the typical negative contraction rate 
operating at small $u$-values eventually wins, making always $u=0$ the only 
stable fixed point.

\section{The spatially extended case}
\label{ext_model}

The past study of stable chaos has revealed that no qualitative difference 
exists, as long as strictly finite chains are considered, with respect to standard 
stable systems. After an exponentially long transient, a periodic behaviour is always attained. 
As it was remarked
in Section II, a similar scenario is indeed exhibited by our stochastic models. 

Eq.~\toref{P0} shows that in the absence of any coupling, there is a finite 
probability $\tilde{P}(u(0))$ for an initial perturbation $u(0)$ to be contracted for an 
infinite amount of time steps and thus of being effectively absorbed. It is worth defining here
\beq 
u_{M}(t) = \max_i{u_i(t)} \quad.
\label{umax} 
\eeq 
In particular, this allows us applying the above reasoning to the coupled 
case, with 
$u_M(0)$ playing the role of $u(0)$; the probability $\tilde{P}$ that an arbitrary, spatially 
extended, perturbation is contracted forever on every site $i$ 
satisfies the inequality
\beq 
\tilde{P}(\{u_i\}_{i=1,\ldots,L} 
\geq \left[ \prod_{n=1}^{\infty} (1-u_{M}(0)a^n)\right]^L = 
\left[\tilde{P}(u_M(0))\right]^L \quad, 
\label{P02} 
\eeq 
where $\tilde{P}(u_M(0))^L$ is small but finite for every finite lattice length $L$. Thus, 
the average time $\tau$
needed for any perturbation to die out (i.e.to be contracted below some 
arbitrarily small thresold value) is finite and does not grow 
faster than exponentially with the system size,
\beq 
\tau \leq  \left[\tilde{P}(u_M(0))\right]^{-L} \quad.
\label{exp2} 
\eeq 
In the active, i.e. chaotic, phase where perturbations propagate with a 
finite velocity, 
$\tau$ grows exactly exponentially (with the system size), since the only 
way for a perturbation to die out is to be contracted in all sites, while in
the inactive phase, the latter is only a sufficent condition and not a
necessary one, and perturbations die out on significantly shorter time
scales.

At variance with the 0-dimensional case, direct numerical simulations of 
the DSM in a 1-d lattice reveal the existence of a regime where finite
perturbations self-sustain. This is clearly shown in Fig.~\ref{tra_cr},
where we have reported the space-time and ensenble average value 
$\tilde{m}$ of the
perturbation for different values of the contraction rate. Above $a=0.60(5)$,
nonzero amplitudes are actually observed. It is, therefore, crucial to
understand the reason why the spatial interactions can stabilize finite
perturbations in spite of the diffusive nature of the coupling.  

\begin{figure}[tcb] 
\centerline{\hbox{\psfig {figure=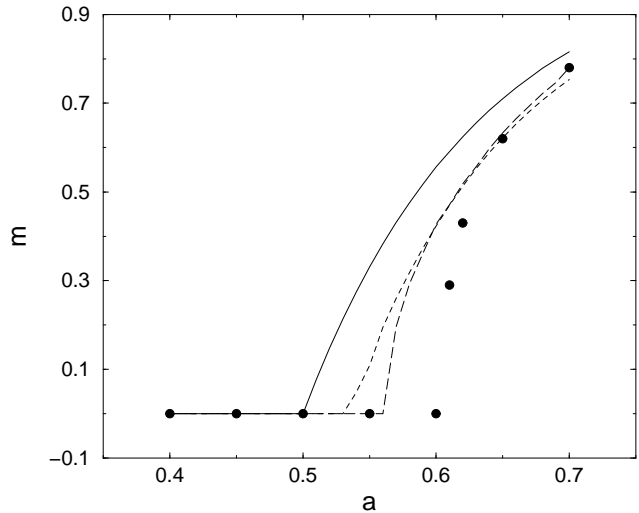,width=7cm,angle=-90}}} 
\caption{The space-time and ensemble average value $\tilde{m}$ of the 
perturbation $u_i(t)$ as a function of the contraction rate $a$. The dots refer 
to the result of numerical
simulations with lattices of size $L=1000$; the solid, dashed, and long-dashed
curves refer to the mean field analysis, the Gaussian approximation
discussed in the appendix B, and the numerical integration of the Frobenius-Perron
equation \toref{paz}, respectively.}
\label{tra_cr} 
\end{figure} 

The natural extension of what we have learnt in zero-dimension consists in 
looking at the joint probability distribution $R(t,u_1,u_2,\ldots,u_n,\ldots)$
over the whole lattice. If the system is sufficiently above 
the transition to stable chaos, it seems reasonable, in a first approximation, 
to neglect spatial correlations. This is certainly incorrect for those
sites that are close to propagating fronts but the fraction of such 
lattice sites is definitely negligible. Accordingly, we approximate the 
joint probability distribution as a product of single-site probabilities $Q(t, u)$, 
\beq
R(t,u_1,u_2,\ldots,u_n,\ldots) \approx Q(t,u_1)Q(t,u_2)\ldots Q(t,u_n)\ldots.
\label{factorize}
\eeq
Within this approximation, the single-site probability distribution corresponding to the 
stochastic dynamics \toref{cont_stoch} satisfies the following 
Frobenius-Perron equation,
\begin{eqnarray}
\nonumber
Q(t+1,u)&=&g(u)\int_0^{\infty}\prod_{i=1}^{N}dv_i\,Q(t,v_i)\,
\delta\left(u-\frac{a}{N}\sum_{i=1}^{N}v_i\right)\\
&&+\,a\Delta\int_0^{\infty}\prod_{i=1}^{N}dv_i\,
Q(t,v_i)\,\delta\left(u-\frac{1}{N\Delta}\sum_{i=1}^{N}v_i\right) 
\nonumber
\end{eqnarray}
\beq
Q(t+1,1)\;=\;\int_{a\,\Delta}^{\infty}du\,u\int_0^{\infty}
\prod_{i=1}^{N}dv_i\,
Q(t,v_i)\,\delta\left(u-\frac{a}{N}\sum_{i=1}^{N}v_i\right),
\;\;\;\;\;\;\;\;\;\;\;\;\;\;\;
\label{paz}
\eeq
where $\delta$ is the Dirac's distribution, $v_i$ is the amplitude of the
perturbation in the $i$th neighbouring site, $N$ is the number of
democratically coupled sites (for later convenience we leave $N$ unspecified -
notice that $N=3$ in a 1-d  lattice with nearest-neighbour coupling) and
\[
g(u)=\left\{
\begin{array}{lr}
1-u & \Delta < u < 1 \\
1-a\Delta & 0 \leq u \leq \Delta
\end{array}
\right..
\]
It is easy to verify that the support of the single-site probability distribution 
$Q(t,u)$ remains confined to the unit interval, provided that this holds true
for the initial condition as well. Furthermore, due to the factorization
hypothesis, the space and ensemble average $m(t)$ defined in Section
II coincides with the simplest ensemble average, i.e. the mean value of
$Q(t, u)$.

\begin{figure}[tcb] 
\centerline{\hbox{\psfig {figure=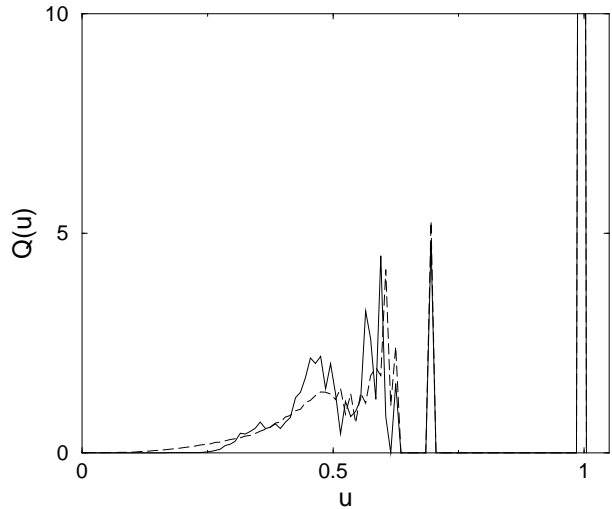,width=7cm,angle=-90}}} 
\caption{Stationary, single-site probability distribution $\tilde{Q}(u)$ 
for $a=0.7$ and $\Delta=0.01$
computed by numerical simulation of the CSM (dashed line) and by numerical
solution of Eq.~\toref{paz} for $N=3$ (solid line). Both
distributions have been obtained subdividing the unit
interval in 100 channels. The $\delta$-peak in $u=1$ has been cut to
magnify all other details.}
\label{fix_p} 
\end{figure} 

In Fig.~\ref{fix_p} we have plotted the single site probability distributions obtained
by directly iterating the stochastic model and the approximate
Frobenius-Perron equation \toref{paz} (solid and dashed line, respectively).
The reasonable overlap confirms the validity of the factorization hypothesis
(at least away from the critical region). The mean value, equal to $0.773$ in former
case, compares with $0.770$ in the latter one, while the variances are 
respectively equal to $0.063$ and $0.061$. Such small differences are due 
to the different behavior of the probability distributions for small values of $u$. 

For $N=1$ (no coupling), Eq.~\toref{paz} corresponds to the 0-dimensional
dynamics discussed in the previous section, and the evolution
equation is exact. It reduces to the linear equation, 
\begin{eqnarray}
Q(t+1,u)&=&g(u)\frac{1}{a}Q(t,\frac{1}{a}u)+a
\Delta^2Q(t,a\Delta u)\nonumber\\
Q(t+1,1)&=&a\int_{\Delta}^{\infty}du\,uQ(t,u),
\label{paz2}
\end{eqnarray}
The only fixed point of this equation is $Q(t,u)=\delta(u)$, i.e. the 
absorbing state. The multifractal spectrum discussed in Sec. III is nothing
but a sophisticate characterization of the convergence towards such a
fixed point.

On the opposite side of the 0-dimensional limit, there is the mean-field 
approximation that corresponds to the limit $N \to \infty$. In this limit, 
statistical fluctuations vanish and the dynamics reduces to the evolution 
of the mean value $m(t)$, that reads
\beq
m(t+1)= \left\{
\begin{array}{lr}
m(t)[2a-a^2m(t)] & \mbox{if}\; m(t) > \Delta\\
m(t)[2a-a^2\Delta] & \mbox{if}\; m(t) \leq \Delta.
\end{array}\right.
\label{meanfield}
\eeq
For $a<a_c=(1-\sqrt{1-\Delta})/\Delta$, Eq.~\toref{meanfield} displays 
the stable fixed point $m_1=0$. This is the same regime found in 0-dimension
and corresponds to the eventual absorption of any initial finite difference. 
Increasing $a$ above $a_c$, the system undergoes a bifurcation: $m_1$ becomes 
unstable and a second (stable) fixed point $m_2=(2a-1)/a^2$ appears. In the 
discontinuous limit  ($\Delta \to 0$) $a_c\rightarrow 1/2$. In
Fig.~\ref{tra_cr} the predictions of the mean-field approach are compared
with the results of direct simulations: we see that, in spite of the
approximations, there are no severe differences and the critical point is 
underestimated by approximately 17\%.

It is rather instructive to notice that the predictions of the
mean-field analysis do coincide with the finite-size Lyapunov exponent
$\Lambda_1$ (for the sake of simplicity, we limit ourselves to consider
the DSM). Since the mean field approximation reduces the CSM dynamics to the 
evolution of a single variable $m(t)$, the comparison can be performed by interpreting 
$\ln[m(t+1)/m(t)]$ as $\Lambda_1$ and recalling that Eq.~\toref{fsle} 
has been derived for the DSM only. Even though Eq.~\toref{fsle} can be obtained
by Eq.~\toref{meanfield}, there is an important difference between
the consequences of the two results. As we discussed in Section III, 
fluctuations keep $u^*=0$ stable for every value of $a$, while in the case of
the mean field analysis, the lack of any fluctuation, due to the formally
infinite number $N$ of neighbours, implies that the fixed point
$u=0$ is truly unstable when $a>1/2$. 

The absence of fluctuations in the mean field limit implies that all
definitions of the finite-size Lyapunov exponent are equivalent. Therefore,
we observe the same scenario previously observed for the standard maximum
Lyapunov exponent: the diffusive coupling shifts the Lyapunov exponent from
the average value of the logarithm of the multiplier (the so-called quenched 
average holding for the single map) towards the logarithm of the average 
multiplier (annealed average, predicted by the mean field analysis).
The important consequence of this shift is that, in the present context,
it can change the stability of the $u=0$ solution leading to the onset of
the chaotic phase.

Anyway, one should not forget that the mean-field anlaysis provides an
approximate solution. For $N$ finite and strictly larger than 1, 
Eq.~\toref{paz} defines a non trivial evolution operator ${\cal Q}$ in a 
functional space. In practice, one can expand the evolution equation into an 
infinite set of equations for, e.g. the momenta $M_k$ of $Q$. An
approximate solution can thus be found by either suitably truncating
the hierarchy of equations or introducing a closure Ansatz. In
App.~\ref{appB} we parametrize the probability distribution as the sum
of a $\delta$-distribution and a Gaussian.  This allows us deriving
three evolution equations for the DSM.

In both cases that we have investigated ($N=2$, 3), we find a scenario similar 
to the one predicted by the mean field analysis. There exists a 
critical value $a_c$ (equal to 0.548 for $N=2$ and to 0.536 for $N=3$) below 
which the dynamics is characterized by the stable fixed point $A=1$, $v=0$, 
$V=0$ (corresponding to the absorbing state $Q(t,v)=\delta(v)$) and above 
which the previous solution becomes unstable, giving rise to a stable 
nontrivial solution. The dependence of $m$ on $a$ reported in Fig.~\ref{tra_cr}
(see the dashed curve in Fig.~\ref{tra_cr}) indicates that the critical
value predicted by this analytic approach improves the mean-field
estimate, but the growth of $m$ above threshold is not as good as one
would like. In fact, there is a qualitative difference with the mean-field 
approach: a further bifurcation (at $a=\bar{a}= 0.855$ for $N=2$ and
$\bar{a} = 0.869$ for $N=3$), where the new 
solution destabilizes too. Such a bifurcation and the slow growth of $\tilde{m}$ 
with $a$ are both consequences of a defect of the approximation: the support 
of the Gaussian extends out of the unit interval. This unphysical property 
becomes increasingly important as soon as the average amplitude of $u$ is
comparable with 1.

Indeed, a better agreement with the direct simulations is obtained by 
iterating numerically the Frobenius-Perron equation (see Fig.~\ref{tra_cr}). 
Neither simulations performed 
with $\Delta = 0$ nor with $\Delta = 0.01$ reveal the second bifurcation
found with the Gaussian approximation, confirming that it is an artifact of
the approximation.
In the continuous case, the bifurcation occurs at the critical value 
$a_c=0.591\ldots$ (for $N=2$) and $a_c=0.568\ldots$ (for $N=3$), to be 
compared with the mean field prediction $a_c=0.501\ldots$.
In the discontinuous case we find $a_c=0.585\ldots$ (for $N=2$) and 
$a_c=0.567\ldots$ (for $N=3$), to be compared with the mean field 
prediction $a_c=0.5$.

From the data reported in Fig.~\ref{tra_cr} for $N=3$ we see that the
factorization hyptohesis reproduces fairly well the behaviour of the
full stochastic model everywhere except for the transition region. This
is not unexpected as it is well known that the correlation length
diverges in the critical region.

\section{Open problems and conclusions}
\label{conclusion}

In this paper we have shown that a simple {\it stochastic} model, specifically
designed to simulate a different response to finite and infinitesimal
perturbations, is able to capture the key features of irregular
behaviour in linearly stable systems. In particular, we have seen that
replacing the sequence of jumps generated by the CML dynamics with a
genuine stochastic process allows for a faithful reconstruction of the
front propagation. The main theoretical advantage of the stochastic
model is the disentanglement between the generation of a pseudo-random
pattern and the evolution of perturbations. In reality the two issues
are interlaced: their separation has allowed us clarifying under which
conditions (finite) perturbations can be effectively sustained throughout an
infinite lattice. In particular, a full consistency exists between the 
CML and the stochastic model in the chaotic regime, since we can state 
that the amplification of finite perturbations contributes to sustain an
irregular regime.

On the other hand, the transition to the ordered phase observed in the
CSM/DSM models does not reproduce the analogous behaviour displayed by
the CML. In this latter case, it was observed that the critical region
is not point-like, but rather extended to a what has been called ``fuzzy
region'', where ordered and chaotic dynamics alternate in a quite
irregular manner \cite{CeLP}. The reason for the difference is that in the CML
model, the absence of local chaos makes the sequence of multipliers
increasingly less random in the transition region. In the stochastic
model, instead, the randomness of multipliers is always assumed a
priori. In spite of such a difference, it is nevertheless instructive to
notice that a transition persists without modifying the stochasticity
in the real space. 

A more precise analogy for the transition investigated in the previous section
is provided by the correspondence with the problem of synchronization in
the presence of external noise. In fact, in this latter context, the
noise represents the (unvaoidable) source of stachasticity in the synchronous
as well as in the asynchronous regime. Since a recent numerical study of the 
synchronization transition in linearly stable system has suggested that it 
belongs to the universality class of Directed Percolation (DP) \cite{BLT},
it is tempting to verify whether the same holds true in our stochastic
models.

As already mentioned, even though the front velocity $v_F$ is a good order 
parameter to characterize the transition, it is quite difficult to obtain a 
reliable estimate of the critical value of the control parameter $a_c$ 
from the vanishing of $v_F$. In fact, finite-size and transient effects 
combined with the existence of wild fluctuations prevent a careful analysis. 
A more efficient method amounts to measuring the dependence of the so-called 
{\sl absorption time} $\tau$ on the system size $L$. This is defined as the 
time required for the space averaged perturbation 
$\bar{u}(t)$ to become smaller than some very 
small, but finite threshold $\Gamma$. Fluctuations of $\tau$ can be efficiently 
reduced by averarging over a sufficiently large ensemble of initial conditions.
In the active phase, $\tau$ is expected to diverge exponentially with
$L$ (stable chaotic regime), while in the absorbing phase, it should
depend at most logarithmically on $L$.
Only at the critical point, $\tau$ exhibits a power law dependence
\beq
\tau(L,a_c) \sim L^z,
\label{sl1}
\eeq
where $z=\nu_l/\nu_{\perp}$ is the so called dynamical exponent. 
We have performed numerical simulations for both the CSM (with $\Delta=0.01$) 
and DSM, averaging over 3000 realizations of the stochastic process and 
over randomly sampled initial conditions. In the CSM we find 
$a_c = 0.6055\ldots$, with $z=1.58\pm 0.02$; in the DSM we obtain 
$a_c=0.6065\ldots$ and $z=1.56\pm 0.06$. The errors have been estimated as 
the maximum deviation from linearity in the log-log plot that has been 
used for extracting the scaling law \toref{sl1} (see, for instance, 
Fig.~\ref{crit_fig}, where $\tau$ has been plotted versus $L$ for 
$\Delta=0.01$ and different values of $a$). These results agree with the most 
accurate numerical estimates of the DP value, $z=1.5807$ \cite{NE}. 

\begin{figure}[tcb] 
\centerline{\hbox{\psfig {figure=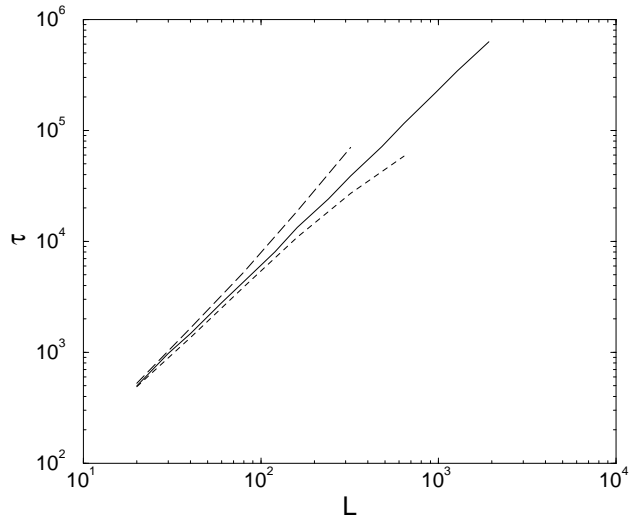,width=7cm,
angle=-90}}}  
\caption{Log-log plot of the absorption time $\tau$ as function of the
system size $L$ in the continuous case ($\Delta=0.01$) near the critical 
value for different control parameter values: $a=0.6051$ (dashed line),
$a=0.655$ (solid line), and $a=0.6061$ (long-dashed line).}
\label{crit_fig} 
\end{figure} 

We also measured the critical exponent $\delta$ associated with the temporal 
decay of the density of active sites $\rho(t)$, i.e. those sites where 
$u_i(t)>\Gamma$: at the critical point $\rho(t)$ is characterized by the 
scaling law
\beq
\rho(t,a_c) \sim t^{-\delta},
\eeq
where $\delta=\beta/\nu_l$ (as usual $\nu_l$, $\nu_{\perp}$ and $\beta$ are 
the critical exponents respectively associated with space and time correlation 
lengths and with the order parameter). By averaging over 3000 realizations
and choosing a sufficiently large value of $L$ to get rid of finite-size
corrections we have found $\delta = 0.150 \pm 0.01$ for the CSM and 
$0.155 \pm 0.005$ for the DSM, to be compared with the DP value 
$\delta=0.1595$ \cite{NE}.

Altogether, our simulations support the hypothesis that the transition belongs
to the same universality class as DP. This may look as an almost trivial 
result, since local spreading is the only mechanism for the propagation of 
perturbations (or, in different languages, active sites, infections). However, 
the whole problem is definitely more subtle, as an absorbing state cannot  
be identified so clearly. In fact, we have already seen that the determination 
of the absorption time requires to fix a somehow arbitrary threshold $\Gamma$, 
and the same is true for the computation of the active sites at a given time. 
Even though we have found that our results are independent of the choice of
$\Gamma$ (provided that it is small enough), this may appear as a numerical 
trick: there is nothing like a true threshold, since no matter how 
small is a perturbation, there is always some finite probability that it 
gives rise to a burst: this is contrary to the existence of a truly absorbing
state. On the other hand, the argument presented in the beginning of the 
previous section to convince the reader that in a finite chain any 
perturbation eventually dies out, confirms the existence of an absorbing 
state: since the smaller is $u_M$ the more likely is that the perturbation
keeps being absorbed, any perturbation $u_i(t)$ has a finite probability to enter
an infinite ``contraction loop'' in which every site $i$ is monotonously contracted
for any time larger than $t$.

It would be nice to put our qualitative arguments on a more rigorous basis,
by definining a suitable finite-size Lyapunov exponent that is negative
below some threshold to indicate that $u^*=0$ is a truly absorbing state.
However, it is not clear whether this could be accomplished, since we know 
that in the chaotic phase, $u^*=0$ should be at the same time ``macroscopically''
unstable since perturbations eventually drive the system towards the only
stable state and ``microscopically'' stable to mean that small enough
perturbations have to be absorbed. In the future, we hope to be able to
clarify whether it is possible to define an indicator that contains both
messages.\\

RL and AP wish to thank R. Kapral for early discussions about a meaningful 
definition of the stochastic model. A. Pikovsky, V. Ahlers and A. Torcini
are acknowledged for useful exchanges of ideas about the characterization of
the transition. A profitable discussion on the 0-dimensional model has been 
carried on with Y. Elskens. Part of this work has been completed thanks to the 
financial support of the NATO contract CRG.973054. We also thank I.S.I. 
in Torino where part of this work was performed.

\appendix
\section{Multifractal distribution}
\label{appA}
In this appendix we report the analytical calculation of the (constant) slope 
of the multifractal distribution \toref{multifractal2} for the zero dimensional
DSM. We shall prove that $H'(\Lambda,u(0))$ (the prime denotes derivative with 
respect to $\Lambda$) is independent of $\Lambda$ and increases monotonously 
with the contraction rate $a$ from -1 (for $a \rightarrow 0$) to 0 
(for $a\rightarrow 1$).\\
First of all, note that the condition $0 \leq n \leq n_0+t $, together with Eq.~\toref{lambdafin},
implies that the support of $H(\Lambda)$ is confined in the interval $[\ln a, 0]$, since
the ratio $\rho = n/t$ can take values 
between 0 and 1 and in the $t \to \infty$ limit  $n_0/t$ vanishes. 

We are now interested in the $\Lambda > \ln a$ case (i.e. the one in which the system received
at least one kick), where, as it was stated in Section III,  
the actual size of the perturbation $u$ 
at time $t$ is unambiguously determined by the time $n < t$ elapsed 
since the last kick (see Eq.~\toref{lambdafin}) and thus, 
for any finite $t$, both $u$ and $\Lambda$ can assume only a discrete set 
of values labelled by $n$. 

Let us now denote the discrete $\Lambda$-derivative of a generic function 
$f(\Lambda)$ as
\beq
D_{\Lambda}(f) = \frac{f(\Lambda+\Delta\Lambda)-g(\Lambda)} {\Delta\Lambda} .
\label{A3}
\eeq 
From Eq.~\toref{multifractal2}, we can approximate the derivative of the 
multifractal distribution $H$ as 
\beq
H'(\Lambda , u(0)) = \lim_{t \rightarrow \infty} \frac{D_{\Lambda}( \ln {\cal P}(\Lambda, t,u(0))}{t},
\label{A4}
\eeq
where $\Delta \Lambda$ is naturally fixed by Eq.~\toref{lambdafin} and 
the discrete character of $n=\ln u / \ln a$, 
\beq
\Delta\Lambda =\Lambda_{\tau+1}(t,u(0))-\Lambda_\tau(t,u(0))=\frac{\ln a}{t}.
\label{A2}
\eeq 
As, for $t \to \infty$, $\Delta \Lambda$ goes 
to $0$, Eq.~\toref{A4} becomes asymptotically exact.

Morover, Eq.~\toref{lambdafin} allows one switching to the $Q$ 
rapresentation of probabilities writing
\beq
H'(\Lambda,u(0)) =  \frac{1}{\ln a} \lim_{t \rightarrow \infty} 
\ln \left[\frac{{\cal P}(\Lambda + \Delta\Lambda, t, u(0))}{{\cal P}(\Lambda, t, u(0))}\right]
= \left. \frac{1}{\ln a} \lim_{t \rightarrow \infty} \ln \left[\frac{Q(t,n+1)}{Q(t,n)}\right]
\right|_{n=n_0+\frac{t\Lambda}{\ln a}}
\label{A6}
\eeq
(remember also that $u(0)=a^{n_0}$). 
Making use of Eqs.~\toref{pippo} and \toref{P0}, we obtain

\beq
H'(\Lambda,u(0)) =  \left. \frac{1}{\ln a} \left[\ln \left(\frac{G(t-n-1,n_0)}{G(t-n,n_0)}\right)
+\ln \left( 1-a^{n+1}\right)\right] \right|_{n=n_0+\frac{t\Lambda}{\ln a}}.
\label{A7}
\eeq

In the limit $t \rightarrow \infty $, the last term in the r.h.s. vanishes,
provided $n$ diverges too (i.e. $\Lambda <0$). We are thus left with the 
following equality,
which holds true for any $\Lambda$ in the open interval $(\ln , 0)$,
\beq
H'(\Lambda,u(0)) = - \frac{\ln \eta(a,n_0)}{\ln a},
\label{A7b}
\eeq
where
\beq
\eta(a,n_0) =\lim_{t \rightarrow \infty} \frac{G(t+1,n_0)}{G(t,n_0)},
\label{A8}
\eeq
the $n$-dependence has been eliminated by shifting the time origin and the 
dependence on the parameter $a$ has been made explicit.
From this equation, we see that the problem of determining the slope of the
multifractal spectrum is equivalent to an eigenvalue problem. In fact, we
can formally write Eq.~\toref{pippo3} as
\beq
\vec{v}(t,n_0) = M \vec{v}(t-1,n_0) ,
\label{A10}
\eeq
where $\vec{v}(t,n_0)$ is the infinite-dimensional vector
\beq
\vec{v}(t,n_0)\equiv(G(t,n_0),G(t-1,n_0),G(t-2,n_0),
\ldots,)
\eeq
and $M$ is a linear infinite-dimensional operator. Therefore, $\eta(a)$ is
nothing but the maximum eigenvalue of the operator $M$. 

From the explicit expression of the recursive equation for $G(t,n_0)$ 
(see Eq.~\toref{pippo3}) 
\beq
G(t,n_0)=G(t-1,n_0)P(0,1)\,a + G(t-2,n_0)P(1,1)\,a^2 + \ldots + 
a^{n_0}\prod_{k=1}^{t-1}(1-a^{k+n_0}) a^{t},
\label{A9}
\eeq
we see that, since both $G(t,n_0)$ and $P(t,1)$ are not larger than 1 (they
are probabilities), the sum in the r.h.s. is bounded from above by the sum of 
the first $t$ powers of $a$. In the limit $t \rightarrow \infty$, the function 
$G(t,n_0)$ converges exponentially fast, so that we are able to truncate 
Eq.~\toref{A9} to order $k$ with an arbitrary precision. This truncation makes
the problem numerically solvable, as the operator $M$ can be approximated
by the finite dimensional matrix $M_k$ (here and below, the subscript $k$ 
stands for $k$th order approximation) that we report here below,
\beq
M_k = \left(
\begin{array}{c}\alpha_1\\1\\0\\0\\\vdots\\0\end{array}
\begin{array}{c}\alpha_2\\0\\1\\0\\\vdots\\0\end{array}
\begin{array}{c}\alpha_3\\0\\0\\1\\\vdots\\0\end{array}
\begin{array}{c}\cdots\\ \cdots\\ \cdots\\ \cdots\\\ddots\\ \cdots\end{array}
\begin{array}{c}\alpha_k\\0\\0\\0\\\vdots\\1\end{array}
\right),
\label{A11}
\eeq
where $\alpha_j=a^j P(j-1,1)$. Numerical estimates of $\eta_k(a)$ indicate that
$k=10$ suffices to attain a good convergence in the whole range of $a$ values
between 0 and 1. For $k=2$ the eigenvalue $\eta_2(a)$ can be computed
analytically yielding,
\beq
H_2'(\Lambda,u(0)) = - 1- \frac{1}{\ln a}\ln\frac{ 1+\sqrt{5-4a} }{2},
\eeq
which goes monotonously from -1 ($a \rightarrow 0$) to $0^-$ 
($a \rightarrow 1$).

\section{An approximate solution of the 1-dimensional problem}
\label{appB}
The Frobenius-Perron operator defined by Eq. \toref{paz} is a functional
equation whose solution requires projecting it onto a finite dimensional
space either via some truncation or a suitable closure hypothesis.
Looking at a typical shape of the stationary probability distribution
in Fig.~\ref{fix_p}, we can see that the multi-peaked structure slows down
the convergence of an expansion in either moments or cumulants.
Accordingly, we have preferred to approximate the probability distribution
as the sum of a Dirac's $\delta$ distribution centered in 1 (the reinjection
point) plus a Gaussian distribution, centered around a point to be determined
self-consistently,  
\beq
Q(t,v) = A_t\delta(v-1) + \frac{1-A_t}{\sqrt{2\pi V_t}}
\exp{\left[-\frac{(v-{\bar v}_t)^2}{2V_t}\right]}.
\label{B1}
\eeq
We can see that $Q(t,v)$ is parametrized by three quantities: $A_t$, the
probability of the $\delta$ component, $V_t$ the variance of the Gaussian,
and ${\bar v}_t$, its average value. 
The mean value $m(t)$ of $Q(t,u)$ is therefore equal to
\beq
m(t)=1-A_t+A_t \bar{v}_t .
\label{boh}
\eeq
It should be noted that our 
approximation is formally ``unphysical'' since the support of any Gaussian 
function is not restricted to the unit interval, but we expect our Ansatz to 
be reasonably correct as long as the probability to be out of the unit 
interval is small enough. 

Entering the above definition of $Q(t,v)$ into Eq.~\toref{paz} and computing
separately the new weight of the $\delta$ component, the average and 
the variance of the
Gaussian, we obtain three evolution equations. In the simplest nontrivial 
case, $N=2$, we have
\begin{eqnarray}
A_{t+1} &=& a\left[ A_t(1-{\bar v}_t)+ {\bar v}_t \right] \nonumber\\
{\bar v}_{t+1} &=& -v_t^2 \left[2 -A_t \right]\frac{a^2}{2} + 
     {\bar v}_t a \left[1-a A_t\right] -
     a\left[1+\frac{a}{2}(V_t-2+A_t)-\frac{1-a}{1-A_t}\right] \\ 
\label{B3}
V_{t+1} &=& v_t^3 \left[ -A_t \left(a-\frac{3}{4}a^2 \right)-
       (1-a)^2 \right] a + {\bar v}_t^2 \left[ A_t( 2 - a) 
       \frac{3}{4}a^2 +a^2-a+\frac{1-a} {1-A_t}\right]+ \nonumber\\
     &&{\bar v}_t \left[ -A_t(1-V_t)\frac{3}{4}a^3 - 
       \frac{3}{2}a^3V_t +a^2(V_t-2) + 2a 
       -\frac{2a(1-a)}{1-A_t} \right] -                \nonumber\\
     &&A_t [a(V_t-1)+2] \frac{a^2}{4} + a^3 + \frac{a^2}{2}
       (V_t-2)+\frac{a^2-a^3}{1-A_t} \nonumber.
\end{eqnarray}
For $N=3$, the case that we have investigated in detail as it corresponds to
a 1-d lattice of democratically coupled maps, the equations read
\begin{eqnarray}
A_{t+1} &=& a\left[ A_t(1-{\bar v}_t)+ {\bar v}_t \right] \nonumber\\
{\bar v}_{t+1} &=& -{\bar v}_t^2[3-A_t]\frac{a^2}{3} + 
 {\bar v}_t(a-\frac{4}{3}a^2A_t) - 
    a\left[ 1+\frac{a}{3}(V_t-3-2A_t)-\frac{1-a}{1-A_t}\right] \nonumber\\
V_{t+1} &=& {\bar v}_t^3 \left[ -\frac{2}{9}A_t^2a^2 +
        \frac{2}{9}A_t (5a-2)a -(1-a)^2 \right] a + \nonumber\\
     &&{\bar v}_t^2 \left[ \frac{2}{3}A_t^2a^3 -
     \frac{2}{3}A_t(2a-3)a^2 + a^2-a+\frac{1-a}{1-A_t}\right] - \\
\label{B4}
   &&\frac{2{\bar v}_t}{3} \left[ A_t^2a^3 - A_t (-3+2a+V_t a)-
      2a\left( -a^2- 1 + \frac{V_t}{2}(2-3a)a - 
      3\frac{1-a}{1-A_t}\right) \right] + \nonumber\\
   &&A_t^2\left(\frac{2}{9}a^3+a^2-a\right) -
     A_t\left[\frac{2}{3}a^3(V_t+\frac{4}{3})-\frac{1}{3}a^2+a\right] + 
     \nonumber \\
  && \frac{8}{9}a^3 +V_t\frac{a^2}{3}-a+ \frac{2(a-a^3)}{1-A_t}.\nonumber
\end{eqnarray}
In spite of the great simplifications involved in the derivation of both sets
of equations, it is still impossible to obtain an analytic expression
even for the critical point $a_c$. The results of the numerical solution of
Eq. \toref{B3} and \toref{B4} are discussed in the text (see Sec.~IV).

\end{document}